# A Fit to the Measurements of the Galactic Cosmic Ray Hydrogen and Helium Spectra at Voyager 1 at Low Energies and Earth Based Measurements at Much Higher Energies Using a Leaky Box Model with Identical Rigidity Independent Source Spectra for the Hydrogen and Helium Nuclei

W.R. Webber[1] and P.R. Higbie[2]

1. New Mexico State University, Department of Astronomy, Las Cruces, NM  88003, USA

2. New Mexico State University, Physics Department, Las Cruces, NM  88003, USA

## ABSTRACT

Voyager 1 data from beyond the heliopause provide the first direct measurements of the interstellar cosmic ray spectra below ~1 GeV/nuc. In this paper we combine these Voyager measurements of H and He nuclei from 3-600 MeV/nuc with higher energy measurements at 1 AU from the BESS and PAMELA experiments up to ~100 GeV/nuc. Using a Weighted Leaky Box Model for propagation in the galaxy, we obtain an excellent fit to these new Voyager observations and the much higher energy spectra up to ~100 GeV/nuc by using source spectra which are $\sim P^{-2.28}$, with the exponent independent of rigidity from low to high rigidities; along with a rigidity dependence of the diffusion path length which is $\sim P^{-0.5}$ at rigidities >1.00 GV, and possibly changing to $\sim P^{1.0}$ at lower rigidities. The nearly identical and, power law independent of rigidity, source spectra that are needed to fit the data on both charges from low to high energies sets limits on the cosmic ray propagation and acceleration processes in the galaxy. This propagation calculation leads to a peak in the differential spectrum of both H and He nuclei at between 20-50 MeV/nuc and an H/He intensity ratio which is nearly constant at a value ~12.0 below ~100 MeV/nuc, both of which are observed in the Voyager data. Solar modulation calculations show that this H/He ratio will be altered in an observable way by any "solar like" modulation effects beyond the heliopause. No modulation like "features" are observed in the H/He ratio thus reinforcing the conclusion that V1 is indeed sampling the local interstellar spectrum.

**Introduction**

The measurement of the hydrogen and helium spectra from 2-600 MeV/nuc beyond the heliopause by Stone, et al., 2013, on Voyager 1 (V1) and the constant intensity of these components of ± a few percent observed now for over 24 months (corresponding to ~7 AU beyond the heliopause), has opened a new window in the study of galactic cosmic rays. Until now the spectra of all galactic cosmic ray nuclei have been obscured below ~1 GeV/nuc by the effects of solar modulation and other more localized heliospheric effects. Above several GeV/nuc these heliospheric effects become more manageable and the high precision measurements of H and He nuclei from BESS (Sanuki, et al., 2000) and PAMELA (Adriani, et al., 2013), have been used to define these spectra up to ~200 GeV/nuc and beyond with quite high accuracy. The window opened by the Voyager measurements below ~1 GV rigidity almost doubles the energy or rigidity range over which the interstellar spectra can now be determined. At lower energies many new parameters enter into both the galactic propagation and the acceleration calculations. Comparing the measured and calculated spectra of both hydrogen and helium components, e.g., the H/He ratio, is particularly valuable below ~100 MeV/nuc, where, at the same energy/nuc, the ionization loss in the interstellar medium is the same for both H and He, whereas the rigidity at the same energy/nuc, perhaps important for both the acceleration and propagation aspects, is a factor ~2 different. Any spectral difference between H and He nuclei is also important with regards to the solar modulation effects at very low modulation levels, near to and just beyond the HP. The lowest energy H and He nuclei and the H/He ratio are especially sensitive to this low level modulation and this will be investigated in this paper.

**The Data**

The data shown in Figures 1 and 2, covering ~5 orders of magnitude in energy, for H and He nuclei respectively, come from several sources; (1) The lowest energies below ~600 MeV/nuc come from V1 as reported by Stone, et al., 2013, and also from V2 in 1998-99 when it was at 60-65 AU (Webber, McDonald and Lukasiak, 2003), at an intermediate level of solar modulation, estimated from the spectra to be ~250 MV (on this scale the modulation at the Earth at the 11 year minimum modulation would be expected to be ~400 MV). (2) The higher energy H and He spectra measured by PAMELA (Adriani, et al., 2013) in 2009 when the cosmic ray intensities at the Earth were at historically high levels, (McDonald, et al., 2010; Mewaldt, et al, 2010) and the solar modulation level at the Earth is estimated to be only 250 MV (Mewaldt, et al., 2010) are shown.  At this level of solar modulation, because of the $P^{1.0}$ dependence of the diffusion coefficient in the heliosphere, the modulation effects become small above a few GV.

Figures 1 and 2 are unconventional.  In the right half of the figures above 1 GeV/nuc, the intensities are multiplied by $E^{2.5}$; in the left half of the figures the intensities are those that are measured.  In effect two separate figures are matched together at 1 GeV/nuc.  This is done because of the steep spectrum at higher energies, which becomes ~$E^{-x}$ or $P^{-x}$ where x is observed to be about -2.78 by the BESS and PAMELA experiments.  As a result of this method of plotting, the vertical scale on the figures is expanded to enable differences of 5-10% in the intensities at both high and low energies to be more easily seen.

An additional advantage of this way of plotting is that it creates two peaks in the overall spectrum of both H and He nuclei which are a factor ~$10^3$ apart in energy; the one between 20-50 MeV and a second apparent higher energy peak at ~20 GeV/nuc where the spectral exponent first exceeds -2.5.  In Figures 1 and 2 the value of measured lower energy peak exceeds the higher energy one by a factor of 4.3 $\pm$ 0.5 for hydrogen and 6.1 $\pm$ 0.5 for helium in the units given.  This difference implies that the energy spectra of the two nuclei may be slightly different.  But note that the calculated energy spectra in our propagation model quite accurately maps this observational data from a few MeV to over 10 GeV/nuc. This model uses identical source rigidity spectra for each component in the propagation calculations.

The black curves in the figures are calculations for source rigidity spectra with exponent ~2.28, along with a diffusion coefficient ~$P^{-0.50}$ above ~1.76 GV with two possibilities for the P dependence of the diffusion coefficient below 1.76 GV.  These calculations will be discussed in the following section.  In these calculations the intensities of all charges are normalized at 10 GeV/nuc in the program.  The charge composition used for these studies is available from the authors.  Our calculation for He includes both $^4$He and $^3$He.  $^3$He contributes as much as 20% of the total He abundance at energies ~1 GeV/nuc.  The calculated curves fit the data for both H and He nuclei well ($\pm$ 10%) at ~50 MeV/nuc and also at 100 GeV/nuc and suggest that the data can be fit by a simple rigidity spectra for each component that has nearly the same exponent between ~0.5 GV to over 200 GV.

Figure 3 shows the H/He ratio that is observed at V1 outside the heliopause along with other measured values of this ratio observed at different modulation levels in the heliosphere. At the lowest energies the H/He ratio and its variation with energy is an extremely sensitive measure of any "localized" solar modulation effects that might influence the spectra. Note that, for zero solar modulation, the propagated LIS H/He ratio is nearly a constant at a value between 11-13 below a few 100 MeV/nuc. This is mainly the result of the interstellar ionization energy loss which dominates the loss processes at these lower energies and indeed may be a factor in the turnover of the LIS spectra at low energies. This energy loss is the same for fully stripped H and He nuclei at the same energy. Since the solar modulation effects are mainly rigidity dependent, the predicted H/He ratio as a function of energy will change rapidly at low energies when solar modulation is introduced, changing from ~12 to ~8 at 10 MeV/nuc for a solar modulation of only 25 MV as shown in the figure (see following sections).

In Figure 3 we also show the effects of a solar modulation of 250 MV on the propagated interstellar H/He ratio. This figure shows the H/He ratio measured by BESS and PAMELA (Sanuki, et al., 2000; Adriani, et al., 2013) and also the results from V2 in 1998 when it was between 60-65 AU (Webber, McDonald and Lukasiak, 2003). The solar modulation at the times of all three measurements is estimated to be between 250 and 400 MV.

### **Diffusive Cosmic Ray Propagation in the Galaxy – The Spectra of H and He Below ~1 GeV/nuc**

In Webber and Higbie, 2009, we derived the LIS H and He and heavier nuclei spectra, which were then used to compare with Voyager measurements out to ~110 AU available at that time. For the interstellar propagation, two models were used, one a modified LBM calculation, and the other a detailed Monte Carlo diffusion calculation. For correspondingly similar propagation parameters the two models gave similar LIS spectra.

It turns out that the LBM calculations in the 2009 paper almost exactly fit the new V1 H and He spectra measured after August 25[th], 2012, reported in Stone, et al., above ~100 MeV/nuc. So for the current paper we have extended these earlier LBM calculations down to the lower limit ~2 MeV/nuc as determined by the limits on the range-energy tables. The cross sections below ~100 MeV/nuc have also been updated (see Webber and Higbie, 2010; Kim, et al., 2002).

The new updated cross section tables, based originally on the formulas in Webber, et al., 2003, above ~100 MeV/nuc but with updated cross sections at lower energies are available by request.

The calculations using the LBM do not contain reacceleration effects, so that the important parameter in this model is the mean escape length, $\lambda$, which depends on the diffusion coefficient and which we take to be, $\lambda = 26.5 \, \beta \, P^{-0.5} \, g/cm^2$. The values in this equation are based on fitting the measured B/C ratio above 0.5 GeV/nuc. This B/C comparison has been augmented by new measurements of this ratio up to 100 GeV/nuc and above by PAMELA (Adriani, et al., 2014) and AMS-2 (Oliva, et al., 2013). Our calculation fits the measured B/C ratio to within 10% $\pm$70% for the individual points over the entire energy range. This comparison is shown in Figure 4. Thus the value of $\lambda$ is determined to be 26.5 $\pm$ 1.5, and the P dependence above ~1 GeV/nuc is 0.50 $\pm$ 0.05.

At lower rigidities the mean path length may change its rigidity dependence depending on the rigidity dependence of the diffusion coefficient. The change in rigidity dependence of the diffusion coefficient is most directly seen by studies using the low energy electrons measured by Voyager (Stone, et al., 2013). From studies of the electron spectrum using both the newer Voyager measurement of electrons at low energies and higher energy electron measurements from PAMELA (Adriani, et al., 2013), Webber and Higbie (2013) determine the change in the exponent between high and low rigidities is between 1.0 and 1.5. The magnitude of this change in exponent is determined by the rigidity $P_0$ at which the break occurs between ~0.316 and 1.0 GV.

Examples of the mean path lengths used in this paper are shown in Figure 5. The $\beta$ dependence of the path length below specific rigidities, $P_0 = 0.56$, 1.00 and 1.76 GV is $P^{3/2}$. Also shown in Figure 5 is the value of $\lambda$ used by Lave, et al., 2013, to interpret the ACE data on $Z \geq 4$ cosmic ray spectra within the framework of the GALPROP propagation model.

Note that Ptuskin, et al., 2006, have argued that a break in the rigidity dependence of the diffusion coefficient should occur at some low rigidity, probably between 1-2 GV, where the GCR energy density approaches that of the confining B field. The change in the exponent of the rigidity dependence is not specified in Ptuskin, et al., although examples are taken with a change ~2.0 or greater in the exponent of the rigidity dependence.

This decrease in the mean path length, $\lambda$, that occurs at lower rigidities is related to the increase in diffusion coefficient at low rigidities (energies) as noted above. As a result the lower rigidity particles more rapidly escape the galaxy. The new Voyager measurements thus probe the features of this diffusion coefficient below ~1 GV, which is 120 MeV/nuc for A/Z = 2 particles. These measurements probe this region down to a few MeV/nuc where the path length may differ by a factor ~5 for different assumptions for the diffusion coefficient at low rigidities. The values of these mean path lengths at 100 MeV/nuc vary from 6-12 g/cm$^2$ and at ~10 MeV/nuc may vary from ~3 g/cm$^2$ to less than 1 g/cm$^2$ depending on the choice of $P_0$ as shown in Figure 5.

For this LBM calculation the IS medium is assumed to be 90% H and 10% He with 15% ionized H and to have an average density of 0.4 cm$^{-3}$. The fraction of ionized H, taken here to be 15%, plays a role in the turnover of the spectra below 20-50 MeV/nuc.

## Discussion – The Spectrum of H and He Nuclei at Both Low and High Energies

This discussion makes use of the data in Figures 1 and 2. The propagation calculations are made for source rigidity spectra ~$P^{2.28}$ with the exponent of the source spectrum independent of rigidity. The resulting energy/nuc spectra, which are what is measured, have energy dependent effects which appear to be due to "propagation"; in essence both the rigidity dependence of the diffusion coefficient as well as the relationship between rigidity and E/nuc spectra influence the shape of the measured E/nuc spectra. At rigidities above $P_0$ the rigidity dependence of the diffusion coefficient is taken to be ~$P^{0.5}$. The source rigidity spectra, which are ~$P^{2.28}$, are thus transformed by the propagation effects and relationships between rigidity and energy/nuc, into E/nuc spectra that become ~$E^{-2.78}$ at high energy, but change spectral shape and exponent at lower energies as seen in Figures 1 and 2. This changing Energy/nuc spectrum is achieved by starting with a simple source rigidity spectra with an exponent independent of rigidity.

For the energy/nuc spectra that are obtained from a constant spectral index of -2.28 in rigidity, the calculated ratio of the intensities of the differential energy/nuclei spectra for H and He nuclei at the peak at 30-50 MeV/nuc and also at the "apparent" peak at 20-30 GeV/nuc (as the spectrum becomes steeper than $E^{-2.5}$), are ~4.0 $\pm$0.5 for H and 5.6$\pm$0.5 for He, consistent with the measured values and their errors quoted earlier.

In other words, the energy/nuc spectra that are measured experimentally look like they are slowly changing with energy but this change is explained by propagation effects on a basic "source" rigidity spectrum which has an unchanging rigidity spectral index of ~2.28 from ~50 MeV/nuc up to ~100 GeV/nuc.

## Discussion – The Significance of the LIS Spectra of H and He at Energies < 100 MeV as Revealed by the Voyager Measurements

This section also makes use of the data in Figures 1 and 2, which includes the peak in the energy/nuc spectra and the decreasing intensities at low energies, and also Figure 3 which shows the measured H/He ratio as a function of energy/nuc.

We start here from the Voyager observations in Figure 3 which show a nearly constant H/He ratio of 12 $\pm$ 1 from a few MeV/nuc up to a few hundred MeV/nuc. This may be unexpected in view of the changes in the H and He energy/nuc spectra as a function of energy that are both observed and calculated. But at about 100 MeV/nuc and below the energy loss by ionization in the IS medium may exceed the diffusion loss from escape from the galaxy as the main loss mechanism for the GCR. In fact, the peaks in the differential energy/nuc spectra at between 20-50 MeV/nuc arise, in part, because of the dominance of this ionization energy loss at still lower energies. The shape and spectral slope at the peak energy and below are determined by this energy loss/nuc which is roughly (~$Z^2/A$) (1/$\beta^2$). This quantity is the same for H and He nuclei at the same E/nuc thus creating a predicted constant intensity ratio.

The constancy of the H/He ratio at low energies that is both predicted and measured makes this ratio a very sensitive indicator of any amount of energy loss that might have occurred as a possible residue of solar modulation beyond the heliopause. This energy loss depends on particle rigidity in the heliospheric modulation models used here (e.g., Gleeson and Axford, 1968).  Since rigidity depends on the factor A/Z, which is 2 times larger for He than H, this modulation process results in a smaller energy loss for He nuclei at the same E/nuc.  The predictions of the changing H/He ratio as a function of the modulation level using modulation potentials =25 MV and 250 MV is shown in Figure 3.  The measurements of the H/He ratio from V2 in 1998 when the spacecraft was between 60-65 AU (Webber, McDonald and Lukasiak, 2003) and the BESS measurements in 1997 (Sanuki, et al., 2000) and the PAMELA measurements in 2007 (Adriani, et al., 2013) are also shown in the Figure.  These BESS and PAMELA measurements help to define a high energy H/He ratio ~17.5 for an energy/nuc spectrum, as well as the solar modulation at the time of the BESS and PAMELA measurements which turns out to be ~350-400 MV.  For the V2 measurements in 1998 the modulation potential is estimated from the data to be ~200-250 MV.  This measurement cannot be extended below ~80 MeV/nuc because of a background of anomalous H and He nuclei.

A residual modulation potential of 25 MeV in this scenario will produce easily observable effects on the H/He ratio at low energies and thus sets a robust limit on any solar "like" modulation process beyond the heliopause.

## Summary and Conclusions

The Voyager measurements of the H and He spectra beyond the heliopause open a new window at low energies for the study of galactic cosmic rays.  In this paper we compare this new spectral data with spectra of these nuclei calculated using a LBM for GCR propagation in the galaxy.  The source spectra in this model are taken to be rigidity spectra with an exponent of ~2.28, which is constant in rigidity throughout the entire range from -0.5 to 200 GV.  The main influence on the shape of this "source" spectrum above 1 GV is from particle diffusion in the galaxy and its rigidity dependence which is described in the LBM in terms of a path length which is ~$P^{-0.5}$ above about 1.0 GV.  At lower rigidities several different path lengths are considered here which have rigidity dependences given by $\lambda \sim \beta^{3/2}$.

We have compared the calculated spectra and intensities with those measured by Voyager.  The shapes of the measured and calculated spectra are very similar, including the observed peaks in the differential spectra at ~20-50 MeV/nuc for both H and He.  To obtain a fit to the V1 data on H and He (as well as electrons) at the lowest energies: (1) The diffusion coefficient needs to have a break about 1 GV and the rigidity dependence of the path length $\lambda$ below this break is taken to be $\sim P^{-1.0-1.5}$. (2) Ionization losses in the interstellar medium begins to influence the spectra at lower energies and are one possible cause of the observed peaks in the differential spectra. Since this energy loss is $\sim Z^2/A$, the H and He spectra will be affected similarly with the result that the predicted H/He ratio is essentially constant at lower energies.

This is what is observed by Voyager 1 (Stone, et al., 2013). This near constancy of the H/He ratio, places limits on possible residual solar modulation effects beyond the heliopause. As an example, a calculation shows that the magnitude of this residual modulation potential must be less than 25 MV, or only 10% of the total solar modulation of 250 MV observed at the Earth in 2009, (Mewaldt, et al., 2010).

The second comparison made in this paper is between the intensities and spectra calculated in the LBM that fit the low energy V1 data and those observed for H and He spectra at higher energies obtained up to ~100 GeV/nuc (~200 GV) by the BESS (Sanuki, et al., 2000) and PAMELA (Adriani, et al, 2013) experiments. This determination of the LIS spectra can now be extended to below a few GV without significant solar modulation effects by using the new Voyager data from ~3-600 MeV/nuc. The comparison of calculated and measured data shows that over this entire range from 3 MeV/nuc to >100 GeV/nuc (0.2-200 GV), both of the observed H and He spectra can be reproduced by a source rigidity spectra $\sim P^{-2.28}$, with the exponent essentially independent of rigidity. This rigidity spectrum is transformed by propagation into the energy/nuc spectra measured by Voyager 1 at lower energies and the BESS/PAMELA spectra measured at higher energies. The uncertainty in the value of this source spectral exponent, as observed from Figures 1 and 2 is $\pm$ 0.04, apart from the uncertainties in the dependence of the diffusion coefficient here taken to be $\sim P^{0.5\pm0.05}$ above ~1 GV.

The H and He spectra below ~100 MeV/nuc are consistent with a break in the rigidity dependence of the diffusion coefficient by ~1.0-1.5 power in the exponent at a rigidity between 0.316 and 1.76 GV.

The remarkable constancy and similarity of the H and He spectral indices in rigidity above 0.5 GV may require a re-examination of the current models for the acceleration and escape of GCR from multiple sources in the galaxy. This includes both the acceleration process and the escape process from the "individual" sources, if indeed the particles we observe represent a collection of thousands of different sources (e.g., SNR).

**Acknowledgments:**  The authors are grateful to the Voyager team that designed and built the CRS experiment with the hope that one day it would measure the galactic spectra of nuclei and electrons.  This includes the present team with Ed Stone as PI, Alan Cummings, Nand Lal and Bryant Heikkila, and to others who are no longer members of the team, F.B. McDonald and R.E. Vogt.  Their prescience will not be forgotten.  This work has been supported throughout the more than 35 years since the launch of the Voyagers by the JPL.

**Figure Captions**

**Figure 1:**  Predictions and measurements of the H nuclei spectra from 3 MeV/nuc to 100 GeV/nuc.  A split figure is used in which the intensities below 1 GeV/nuc are shown on the left hand Y axis, and the intensities x $E^{2.5}$ above 1 GeV/nuc are shown on the right hand axis.  The predictions for the LIS are made using a source rigidity spectra with an index of 2.28 independent of rigidity for both charges and a change in the path length dependence below 1.76 GV as described in Figure 4.  The measurements at higher energies are at different solar modulation levels as described in the text.  The measurements labelled V1 and V2 are after the V1 heliopause crossing on August 25[th], 2012 (Stone, et al., 2013) and from V2 in 1998-99 when this spacecraft was at ~60 AU.  The higher energy measurements are from PAMELA (Adriani, et al., 2013).

**Figure 2:**  This figure is similar to Figure 1 except that the predictions and measurements are for He.

**Figure 3:**  Predictions and measurements of the H/He ratio from 3 MeV/nuc to 100 GeV/nuc.  The curve labelled LIS is the calculated ratio in this paper from the LBM with zero solar modulation.  The data labelled V1 is the H/He ratio measured after the V1 heliopause crossing (Stone, et al., 2013).  The higher energy data are from V2, BESS and PAMELA.  The H/He ratios calculated for residual solar modulation potentials of 25 and 250 MV are also shown.

**Figure 4:**  The B/C ratio as a function of energy.  The measurements as a function of energy are: Engelmann, et al., 1990, orange line; Adriani, et al., 2014, purple line; and, Oliva, et al., 2014, red line.  The predictions of the LBM with dependences of $\lambda \sim P^{-x}$ where x=0.48 and 0.53 are shown.  At lower energies we show two measurements made at a level of solar modulation ~200-250 MV; those of Lave, et al., 2013 (solid circles) and Webber, McDonald and Lukasiak, 2003 (open squares).

**Figure 5:**  The variation of the mean path length, $\lambda$=26.5 $\beta P^{-0.5}$ in g/cm$^2$, in the LBM used in this paper for the propagation of GCR.  All curves have a path length ~$P^{-0.5}$ above 1.76 GV.  The decrease in path length at lower rigidities for the values of $P_0$ = 0.316, 0.56, 1.00 and 1.76 is shown. The solid lines are for a rigidity dependence $\sim\beta^{1.5}$ for the path length at low rigidities as used in this paper. The curve with $P_0=1.76$ most closely follows the dependence used by Lave, et al., 2013, above 1 GV and is used by Lave in the GALPROP program. The Energy/nuc for a A/Z=2.0 particle is indicated along the lower axis.

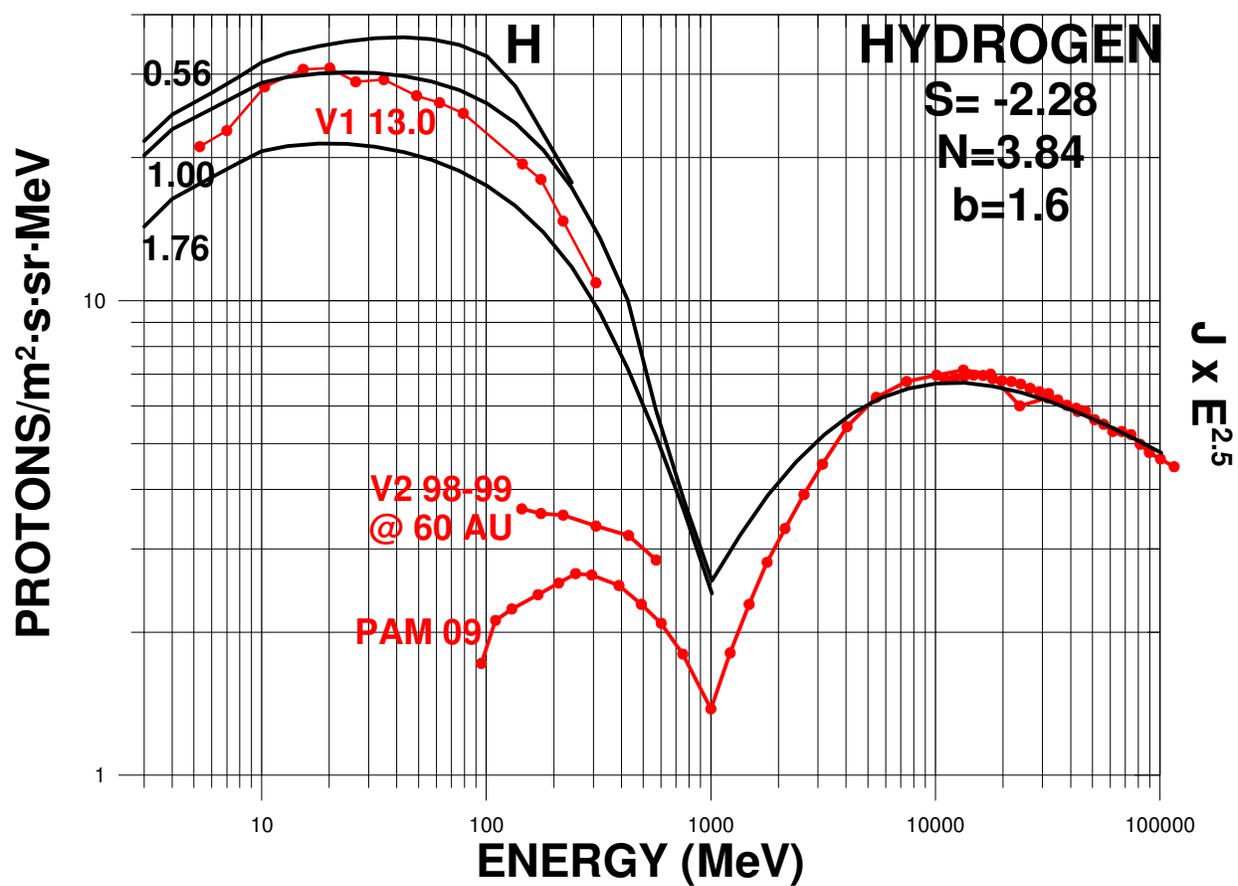

**FIGURE 1**

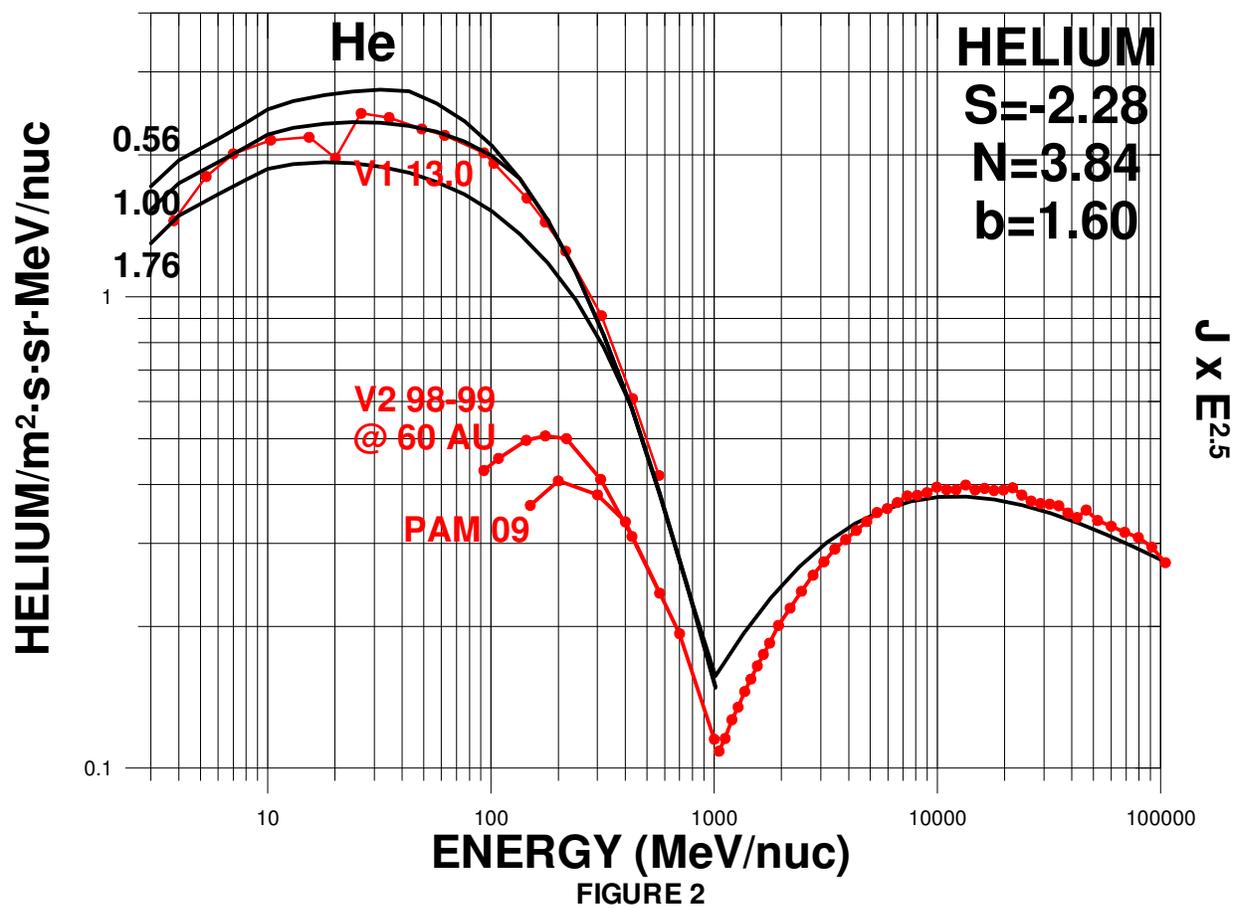

**FIGURE 2**

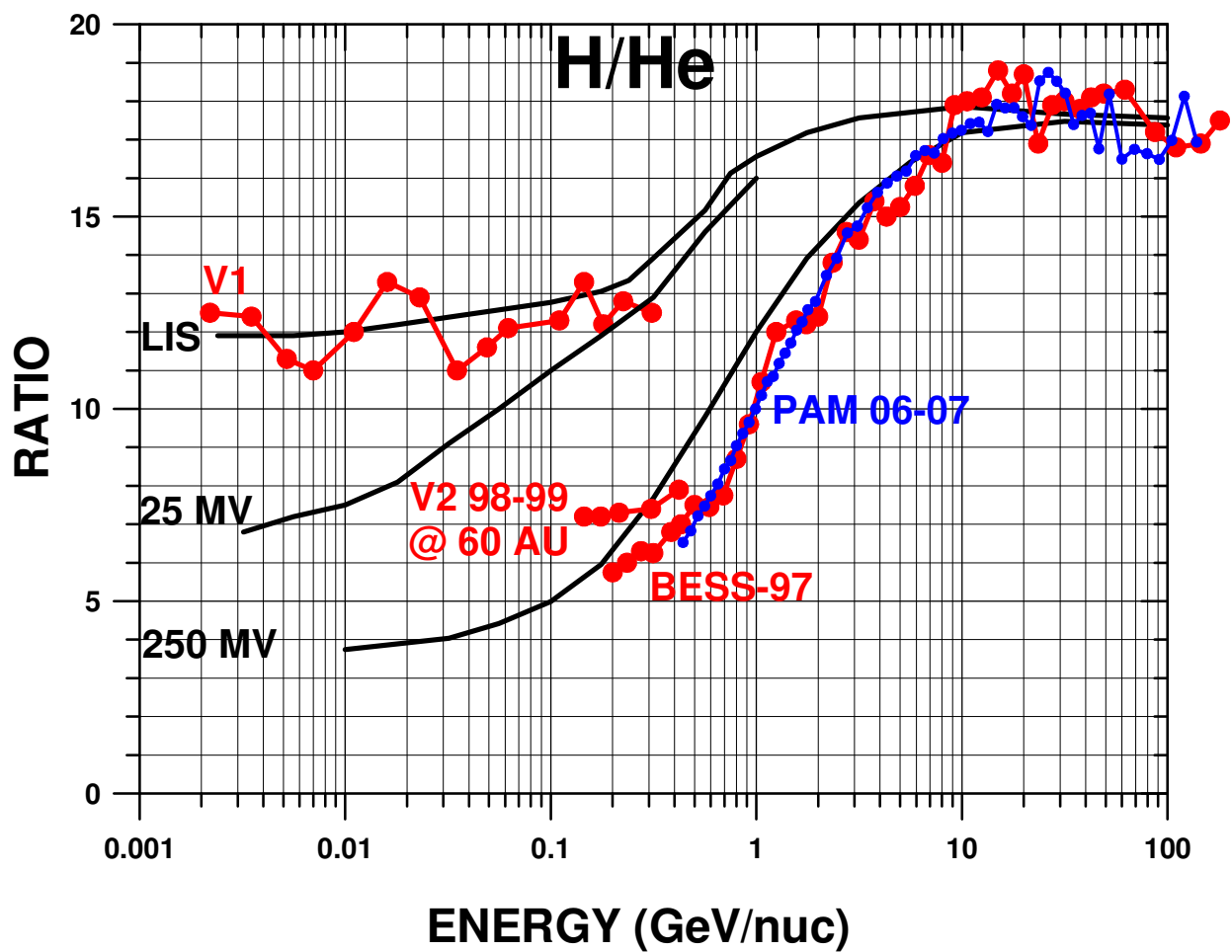

FIGURE 3

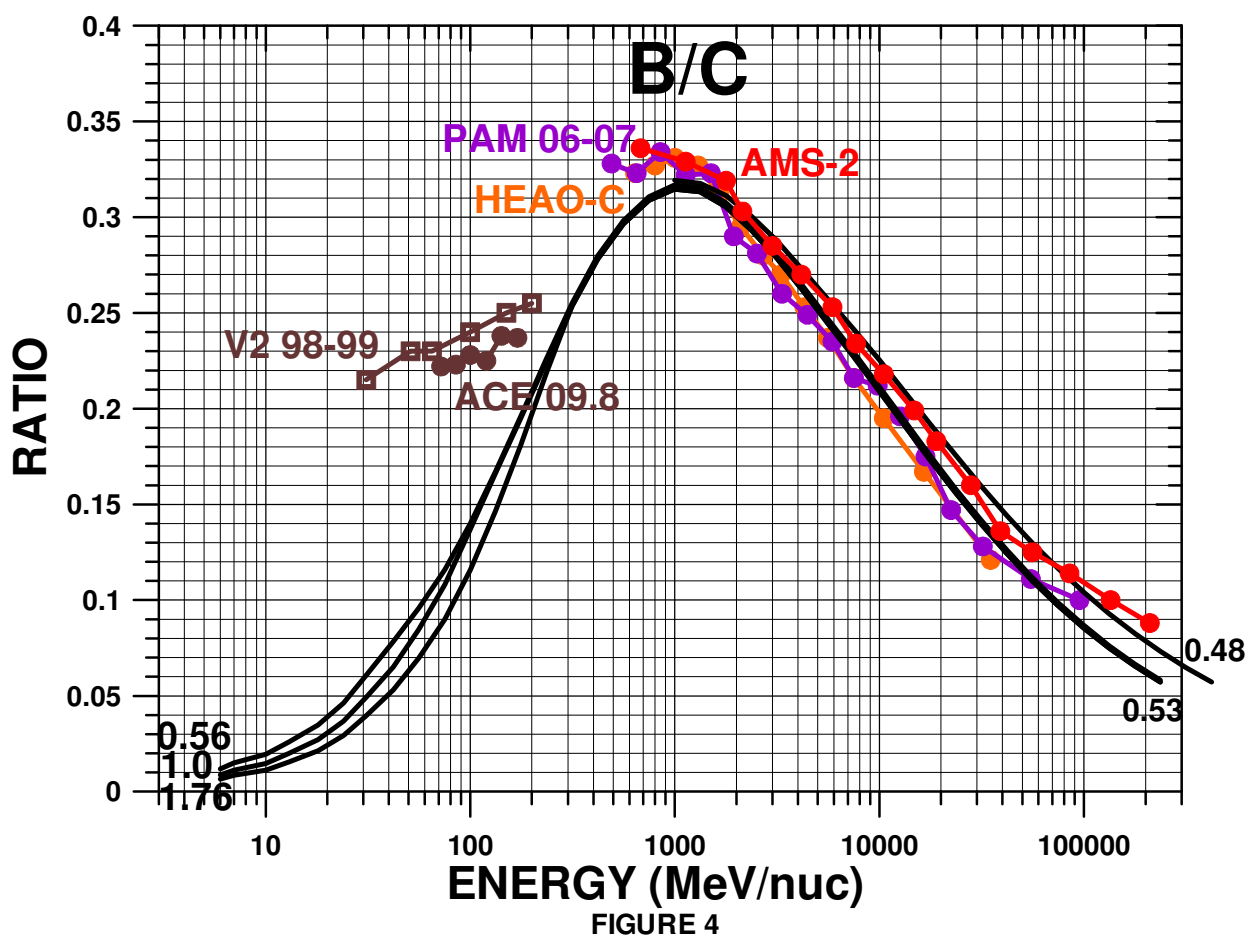

FIGURE 4

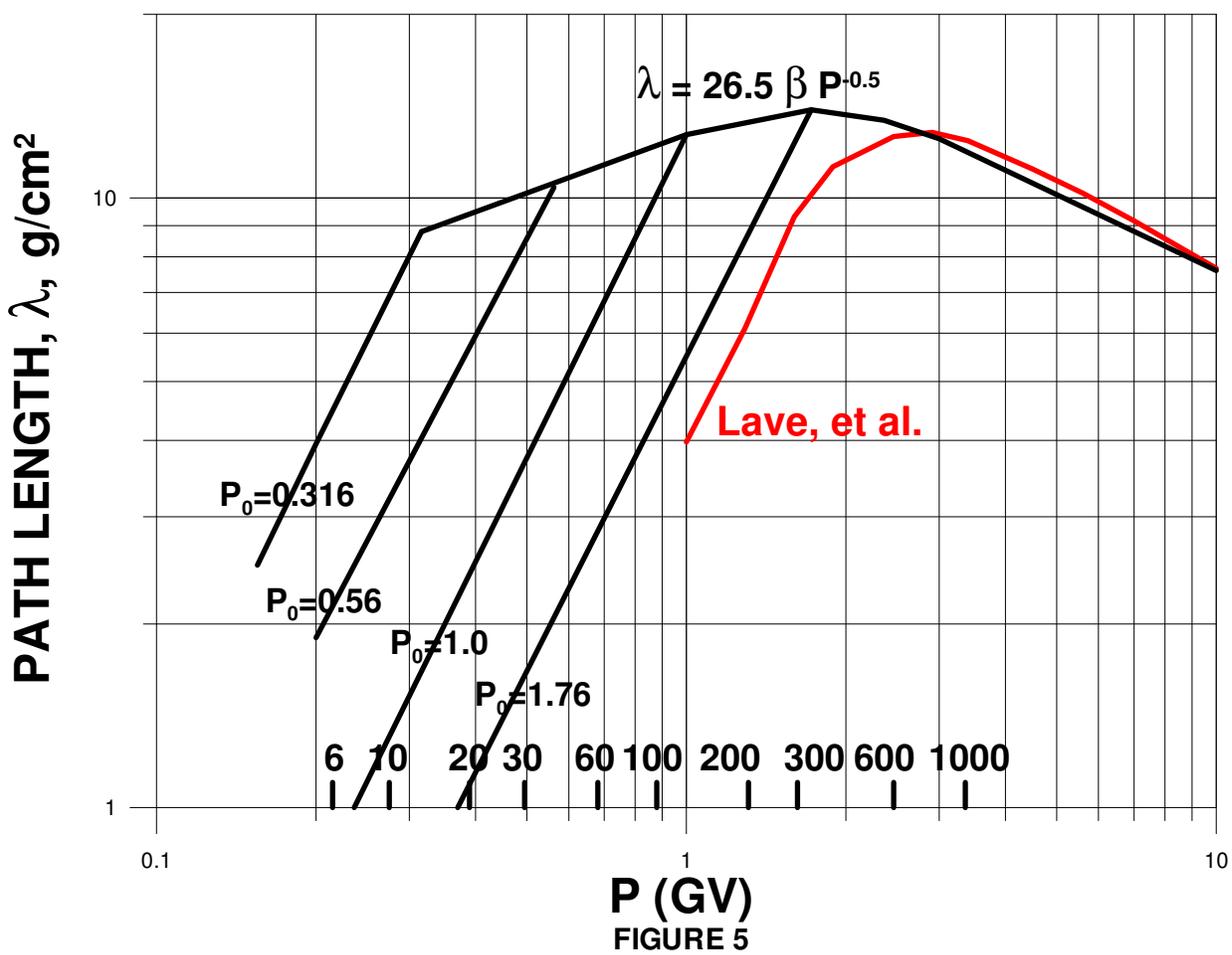

FIGURE 5